# Interactive Cognitive Assessment Tools: A Case Study on Digital Pens for the Clinical Assessment of Dementia

**Daniel Sonntag**
German Research Center for Artificial Intelligence
Stuhlsatzenhausweg 3
66123 Saarbruecken, Germany

**Abstract**

Interactive cognitive assessment tools may be valuable for doctors and therapists to reduce costs and improve quality in healthcare systems. Use cases and scenarios include the assessment of dementia. In this paper, we present our approach to the semi-automatic assessment of dementia. We describe a case study with digital pens for the patients including background, problem description and possible solutions. We conclude with lessons learned when implementing digital tests, and a generalisation for use outside the cognitive impairments field.

## Background

This research is situated within a long-term project (Sonntag 2015) with the ultimate goal of developing cognitive assistance for patients with automatic assessment, monitoring, and compensation in the clinical and non-clinical context. Through the use of mobile devices, multimodal multisensory data (e.g., speech and handwriting) can be collected and evaluated. In the clinical context, we can identify a special target group of interactive cognitive assessment tools as public sector applications: cognitive assistance for doctors in terms of automatically interpreted clinical dementia tests. We think that automatic, and semi-automatic, clinical assessment systems for dementia have great potential and can improve quality care in healthcare systems. Our new project Interakt (Sonntag 2017) with clinical partners from Charité in Berlin complements previous fundamental research projects for non-clinical interfaces for dementia patients (Sonntag 2015; Sonntag 2016) and clinical data intelligence (Sonntag et al. 2016).

In Interakt, we focus on the clinical interpretation of time-stamped stroke data from digital dementia tests. Based on using digital pens in breast imaging for instant knowledge acquisition (Sonntag et al. 2014), where the doctor uses the digital pen for reporting, we now begin to use the digital pen for the patient (Prange et al. 2015). Previous approaches of inferring cognitive status from subtle behaviour in the context of dementia have been made in a clock drawing test, a simple pencil and paper test that has proven useful in helping to diagnose cognitive dysfunction such as Alzheimer's disease. This test is the de facto standard in clinical practice as a screening tool to differentiate normal individuals from those with cognitive impairment and has been digitised in a first version with a digital pen only recently (Davis et al. 2014; Souillard-Mandar et al. 2016). As pointed out in (Davis et al. 2014), the use of (1) a digital pen on paper or (2) a tablet and stylus may distort results by its different ergonomics and its novelty. We implement both interfaces for a selection of standard dementia tests in this case study. This should inform the future development new instances of objective neurocognitive testing methods. In particular, we address the issue of what role automation could play in designing multimodal-multisensor interfaces (Oviatt et al. 2017) to support precise medical assessments. We implemented a set of over 160 signal-level features about the dynamic process of writing such as stroke-level pressure, distance, and duration (Prange, Barz, and Sonntag 2018). This should provide valuable information for conducting (machine learning based) analytics in the context of neurocognitive testing.

## Problem Description

Neurocognitive testing assesses the performance of mental capabilities, including for example, memory and attention. Most cognitive assessments used in medicine today are paper-pencil based. A doctor, physiotherapist or psychologist conducts the assessments. These tests are both expensive and time consuming. In addition, the results can be biased. As a result, we try to understand people, their processes, their needs, their contexts, in order to create scenarios in which Artificial Intelligence (AI) technology can be integrated for digital assessments. We aim to assess and predict the healthcare status with unintrusive sensors such as sensor in digital pens or in tablets. The goal is to improve the diagnostic process of dementia and other forms of cognitive impairments by digitising and digitalising standardised cognitive assessments for dementia. We aim at weekly procedures in day clinics. We base the assessments on clinical test batteries such as CERAD (Morris et al. 1988). We transfer excerpts into the digital world by hand-writing recognition (and sketch recognition) and additional new parameters provided by the digital pen's internal sensors. In this case study we identified that using a digital pen has the following potential benefits:

- the caregiver's time to spend on conducting the test can be reduced;

- the caregiver's time to spend on evaluating the written form can be reduced;
- the caregiver's attention can be shifted from test features while writing (e.g., easy-to-assess completion of input fields) to important verbal test features.
- Digital assessments are potentially more objective than human assessments and can include non-standardised tests and features (for example timing information) whereby previous approaches leave room for different subjective interpretations;
- we can use them to get new features of the pen-based sensor environment, to detect and measure new phenomena by more precise measurement;
- they are relevant for new follow-up checks, they can be conducted and compared in a rigorous and calibrated way;
- we can automatically adapt to intrinsic factors (e.g., sensorimotor deficits) if the user model is taken into account;
- they allow for evidence in the drawing process (e.g., corrections) instead of static drawings that look normal on paper;
- they reduce extrinsic factors (e.g., misinterpreted verbal instructions);
- they can, in the future, be conducted in non-clinical environments and at home.

The challenges we face are three-fold:

1. To identify interface design principles that most effectively support automatic and semi-automatic digital tests for clinical assessments.

2. At the computational level, it is important to investigate approaches to capture both digital pen features and multimodal-multisensor extensions. Some tests assume content features (what is written, language use, perseveration, i.e., the repetition of a particular response such as a word, phrase, or gesture) in usual contexts, as well as para-linguistic features (how is it written, style of writing, pauses, corrections, etc.). These are potential technical difficulties and/or limitations in the interpretation of the results.

3. At the interface level, it is important to devise design principles that can inform the development of innovative multimodal-multisensor interfaces for a variety of patient populations, test contexts, and learning environments.

## Solution

### Interface design principles

Interface design principles are intended to improve the quality of user interface design. According to (Raskin 2000), a computer shall not harm your work or, through inactivity, allow your work to come to harm. According to this guideline, we put much emphasis in developing interfaces that fit well into the scenario, and do not require the patient or the doctor to do more "work" than is strictly necessary. The scenario includes the doctor and the patient at a table in a day clinic (figure 1) which provides most utility and refers to the current paper-pencil based scenario. In the following, we focus on the doctor's assessment task. Here, the term utility refers to whether the doctors' intelligent user interface provides the features they need. We're in the process of evaluating which digital pen features contribute to utility. The conducted cognitive walkthrough started with a task analysis with experts at the clinic that specifies the sequence of steps or actions a doctor requires to accomplish a pencil-paper based assessment task as well as the potential system responses to a digitalised version of it. According to the requirements, we implement a sensor network architecture to observe "states" of the physical world and provide real-time access to the state data for interpretation. In addition, this context-aware application may need access to a timeline of past events (and world states) in terms of context histories for reasoning purposes while classifying the input data. The result of the real-time assessment of the input stroke data and context data is presented to the doctor in real-time, see figure 2. We display (1) summative statistics of test performances, (2) real-time test parameters of the clock drawing test and similar sketch tests, and (3) real-time information about pen features such as tremor and in-air time of the digital pen. These visualisations are based on the set of over 100 signal-level features about the dynamic process of writing.

Usability design choices, how easy and pleasant the interface is to use, are made according to industrial usability guidelines (Sonntag et al. 2010) based on usability inspection methods (Nielsen and Mack 1994) and design heuristics based on the psychophysiology of stress (Moraveji and Soesanto 2012). They can be summarised as follows: For the patient, the digital pen is indistinguishable from a normal pen. So usability is high and (additional) stress is generally low. But the psychophysiology of stress needs to be explored. (Lupien et al. 2007) suggest that some of the "age-related memory impairments" observed in the literature could be partly due to increased stress reactivity in older adults to the environmental context of testing. For the doctor, the psychophysiology of stress needs to be explored, too. There needs to be a possibility to control interruptions (e.g., phone calls) (Moraveji and Soesanto 2012). In general, for both user interfaces, the effects of stress and stress hormones on human cognition are important. (Lupien et al. 2007) enumerate the following stressor characteristics (SC) of interfaces that we use to form further design principles: SC1: Feels unpredictable, uncertain, or unfamiliar in an undesirable manner; SC2: Evokes the perception of losing/lost control. SC3: Has potential to cause harm or loss to one's self or associated objects, living things, or property. SC4: Is perceived as judgment or social evaluative threat including threats to one's identity or self-esteem. Especially SC4 applies in the situation of the patient assessment. Digital pen on normal paper reduces this effect, whereby using a tablet and stylus might increase SC4 stress levels.

### Computational level

The technical architecture is shown in figure 3. As can be seen, at the computational level, there are two intelligent user interfaces, one for the patient (digital pen interaction) and one for the therapist (caregiver interface). The raw pen data is sent to the document processing and indexing server, the pen data processing server provides aggregated pen events in terms of content-based interpretations in RDF (Resource Description Framework). The RDF documents are sent to the data warehouse, together with the RDF meta information. This meta information contains the recognised shapes and text, and text labels, for example. The system attempts to classify each pen stroke and stroke group in a drawing. The second user interface is based on the data warehouse data, and is designed for the practicing clinician. This therapist interface, where the real-time interpretations of the stroke data are available in RDF, is meant to advance existing neuropsychological testing technology according to our interface design principles. Technical details are as follows: First, it provides captured data in real-time (e.g., for a slow-motion playback), and second, it classifies the analysed high-precision information about the filling process, opening up the possibility of detecting and visualising subtle cognitive impairments; also it is zoomable to permit extremely detailed visual examination of the data if needed (as previously exemplified in (Davis et al. 2014)).

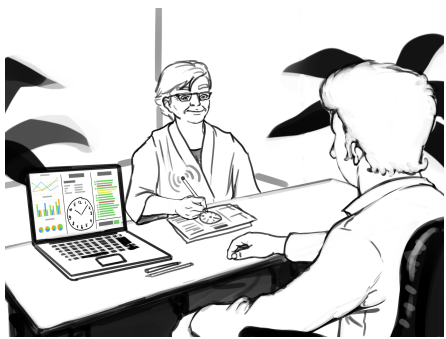

Figure 1: Assessment environment with patient and doctor

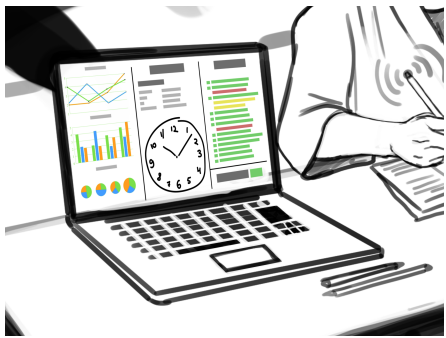

Figure 2: Realtime intelligent user interface for the doctor

Multimodal-multisensor extensions can be implemented with a tablet device (figure 4). Additional modalities can help with the disambiguation of signal- or semantic-level information by using partial information supplied by another modality (Oviatt and Cohen 2015). Additional modalities can help in the analysis of observed user behaviour. When interacting with a tablet computer, multiple built-in sensors can be used in addition.

Besides pen-based input, we consider eye tracking and facial expression analysis via the video signal of the front-facing camera, natural speech captured by the built-in microphone, and additional sensor inputs of modern tablet devices. RGB-based eye tracking is interesting for multimodal interaction with a tablet, because it is deployable using the built-in front-facing camera. However, gaze estimation is erroneous which should be considered in the interaction design (Barz et al. 2018). OpenFace[1] (Baltrusaitis et al. 2018) is an open source toolkit for facial behaviour analysis using the stream of an RGB-webcam. It provides state-of-the-art performance in facial landmark and head pose tracking, as well as facial action unit recognition which can be used to infer emotions. The openSMILE toolkit[2] (Eyben et al. 2013) provides methods for speech-based behaviour analysis and is distributed under an open source license. It offers an API for low-level feature extraction from audio signals and pre-trained classifiers for voice activity detection, speech-segment detection and speech-based emotion recognition in real-time.

### Interface level

The implemented pencil and paper tests are shown in table 1, namely AKT (Gatterer et al. 1989), CDT (Freedman et al. 1994), CERAD (Morris et al. 1988), DemTect (Kalbe et al. 2004), MMSE (Folstein, Folstein, and McHugh 1975), MoCA (Nasreddine, Phillips, and others 2005), ROFC (Canham, Smith, and Tyrrell 2000), and TMT (Reitan 1992).

The pencil and paper tests have been transferred one-to-one, meaning that the digital versions of pen input fields look just as the analog versions. Table 1 shows the absolute percentages of the test questions where the a pen is used to answer them. The selection of the tests accounts for a variety of patient populations and test contexts. Concerning the test context, we can always switch between the digital pen and the tablet and stylus version. The tablet version can always use multimodal-multisensor input to cover additional test contexts.

## Lessons learned

In this section we discuss which choices we have made in the first 18 months of the project, the analysis of alternatives considered, as lessons learned. We focus on specific designs or decision that reduce the potential for failures when considering similar applications.

1. The primary motivation of using a digital pen on normal paper stems from the spatial and temporal precision of the obtained stroke data which provides the basis for an unprecedented degree of precision during analysing this data for small and subtle patterns; classifying the strokes for their meaning is a sketch interpretation task in addition. As a result, we can get assessment data based on what is written or sketched, and how the spatio-temporal pattern looks like. The alternative is using a tablet and stylus turned out to be an additional stress factor for both patients and doctors, as first formative evaluations suggest.

---

[1] https://github.com/TadasBaltrusaitis/OpenFace/
[2] https://audeering.com/technology/opensmile/

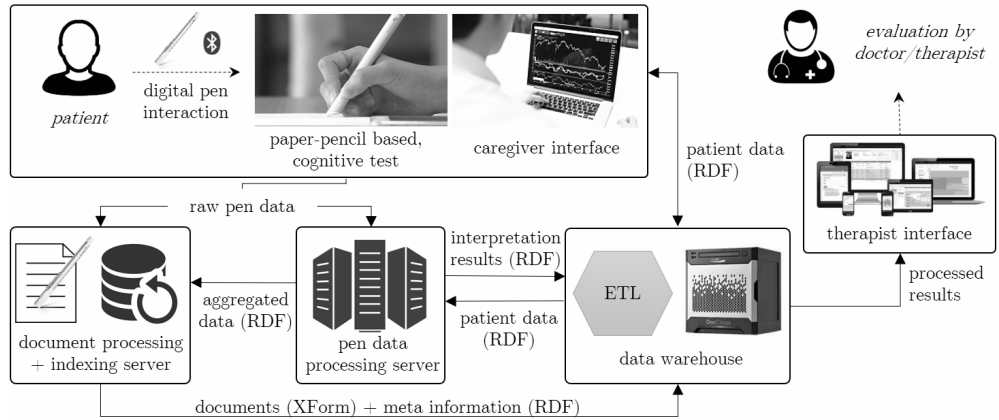

Figure 3: Architecture

| name | approx. time needed | pen input | symbols |
|---|---|---|---|
| AKT - Age-Concentration | 15 min | 100% | cross-out |
| CDT - Clock Drawing Test | 2-5 min | 100% | clock, digits, lines |
| CERAD - Neuropsychological Battery | 30-45 min | 20% | pentagrams, circle, diamond, rectangles, cubes |
| DemTect - Dementia Detection | 6-8 min | 20% | numbers, words |
| MMSE - Mini-Mental State Examination | 5-10 min | 9% | pentagrams |
| MoCA - Montreal Cognitive Assessment | 10 min | 17% | clock, digits, lines |
| ROCF - Rey-Osterrieth | 15 min | 100% | circles, rectangles, triangles, lines |
| TMT - Trail Making Test | 3-5 min | 100% | lines |

Table 1: Comparison of the most widely used cognitive assessments

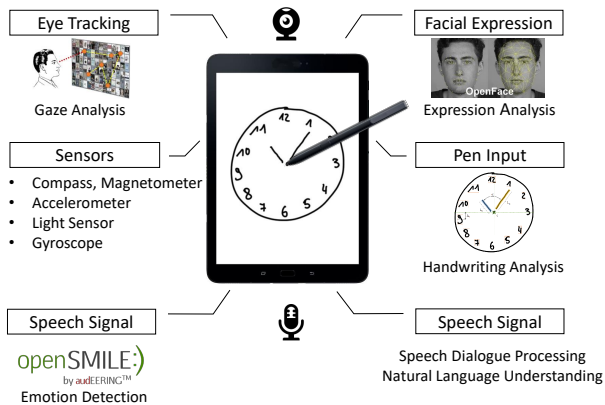

Figure 4: Multimodal-multisensor tablet device

As a result, the formative evaluations with patients will be done on the digital paper version. This choice restricts the possibility to gather multimodal data from a tablet, which provides the same spatial and temporal precision of the obtained stroke data.

2. While the tablet version is not always the first choice, the technical implementation is much easier than the digital pen on normal paper version. The reason is the complicated SDK for creating the digital paper forms on normal paper.

3. How we would go about collecting data and test the questions we are examining raise some issues: How will the data from the experiment be gathered without violating privacy regulations? For example, video capture is not possible. Will it be complete? We need a method to capture assessment results (or corrections/comments) from the doctor while he or she is using the doctor's interface. Will it interfere with the anticipated normal use? Here, the enumeration of the stressor characteristics need to be completed and turned into interface design principles.

4. A version for self-assessment at home for the patient needs to have an ability to control interruptions (e.g., phone calls) (Moraveji and Soesanto 2012).

5. The digitalisation of widely used cognitive assessments has four consecutive steps: first, the one-to-one transfer from a paper and pencil test to a digital version; second, the selection of pen features that are relevant for the classification task: third, the adaptation of the caregivers' instructions to include automatically interpreted test results. And fourth, the inclusion of multimodality and multisensor data for additional test parameters.

6. Digital assessments allow for evidence in the drawing process (e.g., corrections) instead of static drawings that look normal on paper. Doctors need to be instructed when to use the "replay" function. To propose replaying a writing scene for further inspection is another interesting classification task of the system-initiative interface.

7. The coverage of implemented tests is rather independent of the availability of suitable patient populations and test subjects. It is rather difficult to get the critical amount of conducted tests for machine learning experiments to find subtle pattern that are sensitive or specific to dementia assessment.

## Generalisation and Future Research

Using digital pens for the assessment of dementia can be generalised in several ways, most notably for use by those in the cognitive impairments field. Digitalised dementia test can be used for the detection of other neurodegenerative diseases such as Parkinson. Some of the described tests in table 1 have already been used in this direction. In addition, this work could help returning veterans suffering from TBI (traumatic brain injuries). While acute TBI can be life threatening, TBI also can have long-term sequelae including cognitive and physical disability, post-concussion syndrome (PCS), and may contribute to the development of chronic traumatic encephalopathy (CTE). (J Wagner et al. 2011) used the CDT to assess cognition and predict inpatient rehabilitation outcomes among persons with TBI. Doctors working in inpatient neurorehabilitation settings are often asked to evaluate the cognitive status of persons with TBI and to give opinions on likely rehabilitation outcomes. In this clinical setting, several other digital pen tests could be used for cognitive assessment and outcome predictor among inpatients receiving neurorehabilitation after TBI. It should be possible to better monitor the rehabilitation outcome. As explained above, digital assessments could be relevant for new follow-up checks, they can be conducted and compared in a rigorous and calibrated way.

Future research in the clinical domain includes pen-based assessments to treat patients in an automatic fashion and from multimodal input. For interpreting verbal utterances of the CERAD test battery for example (therapists have problems in taking notes of user answers and comments while conducting a test), a dialogue framework can be used in the future (Neßelrath 2016). Combining speech and pen input (active input) should, in the future, be explored towards multimodal approaches to determining cognitive status. This can be done through the detection and analysis of subtle behaviours and skin conductance sensors. In addition, research in the multimodal-multisensor domain investigate observable differences in the communicative behaviour of patients with specific psychological disorders (DeVault et al. 2014), and detection of depression from facial actions and vocal prosody (Cohn et al. 2009). With a selection of pen-based tests for those disorders, a combined analysis could be made. Another direction is to include the digital pen analysis into email apps where you can use handwriting as the process of writing combines several cognitive and motor-functions that can be assessed outside the scope of standard cognitive test batteries. This means the basic functionality can be turned into an application on a smartphone with a stylus. Given the ubiquity of smartphones with cameras, multimodal-multisensor features based on camera, sensor and the speech signal could be explored in field studies including quantitative dimensions.

Using digital pens for the assessment of dementia can be generalised for use by those outside the cognitive impairments field. Generalisations can for example be implemented in the educational context, in undergraduate and graduate student populations: Student emotional health is at an all-time low as students face increasing stress and academic pressures (Kelley, Lee, and Wilcox 2017). Stress can lead to negative psychological and physical effects over time (Konrad et al. 2015; Moraveji and Soesanto 2012). We could focus on mental wellness on the issues, both clinical and non-clinical, of stress, anxiety, and depression, as they are the three mental health concerns most prominent for students.[3] Researchers have begun to draw correlations between tracked behaviours and self-reported indicators of mental well-being. A digital pen is an ideal tool to monitor a student's writing behaviour. With digital pens, the indicators of stress, anxiety and depression could be learned and monitored in the general student population. Earlier work on kinematic analysis of handwriting movements with a tablet and pressure-sensitive stylus suggest that specific motor dysfunction when writing is an indicator of depression (Schröter et al. 2003). Likewise, the design of future educational interfaces may include digital pen analysis for more expressively rich and flexible communication interfaces that can actually stimulate human cognition (Oviatt 2013).Current research investigates the use of handwriting signal features to predict domain expertise in several educational contexts (Oviatt et al. 2018). The trend towards multimodal learning analytics becomes apparent, where natural communication modalities like writing (or speech) are complemented with gestures, facial expressions, and physical activity patterns. The combination of our low-level stroke features, with selected components of the implemented cognitive tests, together with the domain expertise domain prediction task in (Oviatt et al. 2018) might open up opportunities to design new educational technologies based on individualised writing data for user modelling.

In addition, connecting to education might broaden the application of interactive cognitive assessment tools in future research. (Forbus et al. 2008) describe CogSketch, an open-domain sketch understanding tool for education. The 2018 version includes an authoring tool for an automatic sketch understanding system. One part of that is modelling the semantics of visual and spatial properties in a human-like way. The clock drawing test can be seen as an instance of a visual and spatial sketch understanding task. It should be interesting to use CogSketch to implement digital versions of cognitive assessment tasks, i.e., the semantic specification of the tasks, based on our interpretation of the symbols drawn with the digital pen (see table 1, right). Experiments similar to digital dementia tests could enable us to model spatial skills and learning processes.

---

[3]https://sites.psu.edu/ccmh/files/2018/01/2017_CCMH_Report-1r3iri4.pdf

## Acknowledgements

This research has been funded by the Federal Ministry of Education and Research (BMBF) under grant number 16SV7768. See the DFKI page http://www.dfki.de/MedicalCPS/?page_id=725 for more information.